\documentclass[12pt]{article}
\usepackage{amsfonts}
\usepackage{amssymb}

\def\hybrid{\topmargin -20pt    \oddsidemargin 0pt
        \headheight 0pt \headsep 0pt
        \textwidth 6.25in       
        \textheight 9.5in       
        \marginparwidth .875in
        \parskip 5pt plus 1pt   \jot = 1.5ex}

\hybrid

\def\baselinestretch{1.2}

\catcode`\@=11

\def\marginnote#1{}
\newcount\hour
\newcount\minute
\newtoks\amorpm
\hour=\time\divide\hour by60
\minute=\time{\multiply\hour by60 \global\advance\minute by-\hour}
\edef\standardtime{{\ifnum\hour<12 \global\amorpm={am}%
        \else\global\amorpm={pm}\advance\hour by-12 \fi
        \ifnum\hour=0 \hour=12 \fi
        \number\hour:\ifnum\minute<10 0\fi\number\minute\the\amorpm}}
\edef\militarytime{\number\hour:\ifnum\minute<10 0\fi\number\minute}

\def\draftlabel#1{{\@bsphack\if@filesw {\let\thepage\relax
   \xdef\@gtempa{\write\@auxout{\string
      \newlabel{#1}{{\@currentlabel}{\thepage}}}}}\@gtempa
   \if@nobreak \ifvmode\nobreak\fi\fi\fi\@esphack}
        \gdef\@eqnlabel{#1}}
\def\@eqnlabel{}
\def\@vacuum{}
\def\draftmarginnote#1{\marginpar{\raggedright\scriptsize\tt#1}}

\def\draft{\oddsidemargin -.5truein
        \def\@oddfoot{\sl preliminary draft \hfil
        \rm\thepage\hfil\sl\today\quad\militarytime}
        \let\@evenfoot\@oddfoot \overfullrule 3pt
        \let\label=\draftlabel
        \let\marginnote=\draftmarginnote
   \def\@eqnnum{(\theequation)\rlap{\kern\marginparsep\tt\@eqnlabel}%
\global\let\@eqnlabel\@vacuum}  }

\def\preprint{\twocolumn\sloppy\flushbottom\parindent 2em
        \leftmargini 2em\leftmarginv .5em\leftmarginvi .5em
        \oddsidemargin -.5in    \evensidemargin -.5in
        \columnsep .4in \footheight 0pt
        \textwidth 10.in        \topmargin  -.4in
        \headheight 12pt \topskip .4in
        \textheight 6.9in \footskip 0pt
        \def\@oddhead{\thepage\hfil\addtocounter{page}{1}\thepage}
        \let\@evenhead\@oddhead \def\@oddfoot{} \def\@evenfoot{} }

\def\numberbysection{\@addtoreset{equation}{section}
        \def\theequation{\thesection.\arabic{equation}}}

\def\underline#1{\relax\ifmmode\@@underline#1\else
        $\@@underline{\hbox{#1}}$\relax\fi}

\def\titlepage{\@restonecolfalse\if@twocolumn\@restonecoltrue\onecolumn
     \else \newpage \fi \thispagestyle{empty}\c@page\z@
        \def\thefootnote{\fnsymbol{footnote}} }

\def\endtitlepage{\if@restonecol\twocolumn \else \newpage \fi
        \def\thefootnote{\arabic{footnote}}
        \setcounter{footnote}{0}}  

\catcode`@=12
\relax

\def\figcap{\section*{Figure Captions\markboth
        {FIGURECAPTIONS}{FIGURECAPTIONS}}\list
        {Figure \arabic{enumi}:\hfill}{\settowidth\labelwidth{Figure
999:}
        \leftmargin\labelwidth
        \advance\leftmargin\labelsep\usecounter{enumi}}}
 \relax
\def\tablecap{\section*{Table Captions\markboth
        {TABLECAPTIONS}{TABLECAPTIONS}}\list
        {Table \arabic{enumi}:\hfill}{\settowidth\labelwidth{Table
999:}
        \leftmargin\labelwidth
        \advance\leftmargin\labelsep\usecounter{enumi}}}
 \relax
\def\reflist{\section*{References\markboth
        {REFLIST}{REFLIST}}\list
        {[\arabic{enumi}]\hfill}{\settowidth\labelwidth{[999]}
        \leftmargin\labelwidth
        \advance\leftmargin\labelsep\usecounter{enumi}}}
 \relax
\makeatletter
\newcounter{pubctr}
\def\publist{\@ifnextchar[{\@publist}{\@@publist}}
\def\@publist[#1]{\list
        {[\arabic{pubctr}]\hfill}{\settowidth\labelwidth{[999]}
        \leftmargin\labelwidth
        \advance\leftmargin\labelsep
        \@nmbrlisttrue\def\@listctr{pubctr}
        \setcounter{pubctr}{#1}\addtocounter{pubctr}{-1}}}
\def\@@publist{\list
        {[\arabic{pubctr}]\hfill}{\settowidth\labelwidth{[999]}
        \leftmargin\labelwidth
        \advance\leftmargin\labelsep
        \@nmbrlisttrue\def\@listctr{pubctr}}}
 \relax
\makeatother
\newskip\humongous \humongous=0pt plus 1000pt minus 1000pt

\newif\ifdtup

\relax

\def\be{\begin{equation}}
\def\ee{\end{equation}}
\def\ba{\begin{eqnarray}}
\def\ea{\end{eqnarray}}

\def\del{\partial}

\def\G{\Gamma}
\def\d{\delta}

\def\m{\mu}
\def\n{\nu}

\def\L{\Lambda}
\def\s{\sigma}
\def\S{\Sigma}

\def\cL{{\cal L}}

\def\bs{\bigskip}
\def\no{\noindent}

\def\qq{\qquad}
\def\bl{\bigl}
\def\br{\bigr}

\def\IR{\relax{\rm I\kern-.18em R}}

\def \ha {{1\over 2}}

\def \ov {\over}

\def\IR{\relax{\rm I\kern-.18em R}}
\def\inv{^{\raise.15ex\hbox{${\scriptscriptstyle -}$}\kern-.05em 1}}

\def\cL{{\cal L}}

\def\tr{{\rm tr}}

\begin{document}


\renewcommand{\theequation}{\arabic{equation}}

\newcommand{\beq}{\begin{equation}}
\newcommand{\eeq}[1]{\label{#1}\end{equation}}
\newcommand{\ber}{\begin{eqnarray}}
\newcommand{\eer}[1]{\label{#1}\end{eqnarray}}
\newcommand{\eqn}[1]{(\ref{#1})}
\begin{titlepage}
\begin{center}

\hfill NEIP-01-005\\
\vskip -.1 cm
\hfill hep--th/0106126\\
\vskip -.1 cm
\hfill May 2001

\vskip .5in

{\large \bf Dynamical emergence of extra dimensions and \break
warped geometries}

\vskip 0.4in

{\bf Konstadinos Sfetsos}
\vskip 0.1in
{
Institut de Physique, Universit\'e de Neuch\^atel\\
Breguet 1, CH-2000 Neuch\^atel, Switzerland\\
\footnotesize{\tt sfetsos@mail.cern.ch}}\\

\end{center}

\vskip .3in

\centerline{\bf Abstract}

\noindent
We present four-dimensional gauge theories in Minkowski spacetime
which effectively generate in certain energy regimes five-dimensional
warped geometries whereas, in general, the fifth dimension is latticized.
After discussing in detail several general aspects in such theories we present
a number of exactly solvable examples.
We also point out how a particular case,
defined in an $N$-sided polygon and having a $Z_N$ symmetry, has a similar
realization in an appropriate supersymmetric setting
with D3-branes.

\vskip .4in
\noindent

\end{titlepage}
\vfill
\eject

\def\baselinestretch{1.2}
\baselineskip 16 pt
\noindent

\section{Introduction}

A quite attractive idea that has received a lot of attention is that our
spacetime is higher than four-dimensional. This idea resonates well with
string theory considerations,
as well as with the more recent realization, in connection with a geometrical
reformulation of the hierarchy problem, that such
extra dimensions can be even macroscopic in size since this is allowed by the
present day experiment evidence \cite{ADV}.
In most of the literature extra dimensions is an assumption that one makes.
Their number, size and other properties are constrained from
theoretical consistency and experimental results.
In such approaches the important point is that, certain phenomena
in our four-dimensional world are just ``projections'' of Physics
in higher dimensions, where the fundamental theories are defined.

From a philosophical point of view it is quite different to demand that
extra dimensions emerge dynamically and that the
resulting theories are only effective descriptions of
fundamental four-dimensional theories in appropriate energy regimes.
Generically, such extra dimensions should emerge between
certain energy scales, but the corresponding theories should look
four-dimensional at very high energies in the ultraviolet (UV) and very
low energies in the infrared (IR). That guarantees the four-dimensional
nature of interactions at a macroscopic as well as at
a microscopic level, provided the four-dimensional theories are well
behaved, i.e. are renormalizable, asymptotically free etc.

In studies of gauge theories at strong coupling using the AdS/CFT
correspondence, we have constructed
a model which shares the desired features we just mentioned \cite{sfe1}.
In particular, we constructed a D3-brane solution describing
part of the Coulomb branch
of ${\cal N}=4$ SYM theory with gauge group $SU(Nk)$, in which
vacuum expectation values (vev's) to two of the six scalars
transforming in the adjoint of the gauge group were given.
The scalar vev's were taken to lie in an $N$-sided polygon and in this way the
original $SO(6)$ global invariance of the theory breaks to an
$SO(4)\times Z_N$.
This theory turns out to look four-dimensional for
very large and very low energies and effectively five-dimensional for a
wide range of energies in between. This conclusion was also verified by
examining the potential energy for a quark-antiquark pair in \cite{brand1}.
This potential, as we will also see in this paper,
has the usual Coulombic bahaviour $1/r$ for very large and
very low distances, but it behaves as $1/r^2$ in between, hence
confirming the effective emergence of a fifth spatial dimension.

More recently a similar idea was proposed in \cite{ACG1} and in
\cite{HPW1} directly in a conventional field theory which need not be
supersymmetric. A four-dimensional field theory was
defined in an $N$-sided polygon
with gauge group $SU(N)\times SU(k)$. Fermions provide the necessary link
and transform bilinearly under nearest neighbor pairs of gauge
transformations. If the dynamical energy scale for the $SU(N)$ gauge
theory $\L_N$ is much larger than that for the $SU(k)$ theory $\L_k$,
then for energies comparable to $\L_N$ the fermions condense in pairs
since at these scales each of the $SU(N)$ gauge groups becomes strong,
whereas the gauge groups $SU(k)$ can still be treated perturbatively.
The authors of \cite{ACG1} and \cite{HPW1} argued that
the appropriate theoretical description is in terms of a
latticized gauge theory for $SU(k)$
interacting with a $\s$-model that provides the link between the different
sites with nearest neighbor type of interaction.
The gauge group is broken down to the diagonal subgroup $SU(k)$ with the
familiar Higgs mechanism. The vev's
$v$ were chosen to be the same at each site and are inversely proportional
to the lattice spacing. It was subsequently shown that the
theory looks four-dimensional for distances smaller
than the lattice spacing $1/v$ and for distances
larger than the size of the polygon $\sim N/v$. However, in between,
for $1/v\ll r\ll N/v$
the theory looks five-dimensional with the fifth dimension
builded up by the Kaluza--Klein states of the large circle of radius $N/v$.
The situation is the same as in \cite{sfe1} in the sense
that in the latter case the $\s$-model is provided by
the six scalars in the ${\cal N}=4$ SYM
theory that transform in the adjoint of the gauge group.
Also in both cases the scalars assume vev's precisely the same way.
This will be shown explicitly in section 4.

We note that a similar idea of a latticized array of gauge theories
has been proposed in the past in an attempt to explain
the hierarchy of strengths between the electroweak and strong
interactions \cite{Halpern}. Also, there is a close relation
to lattice gauge theories with anisotropic couplings
in five or higher dimensions, which, for certain values of the couplings,
exhibit four-dimensional layer-type
phases as it was proposed in \cite{FuNi} and further elaborated on, in
\cite{FuNi2,Stam}.
We also note recent related work that has appeared,
with emphasis into more
phenomenological applications in \cite{CTPW2}-\cite{Kim}
and some connection with non-commutative geometry in \cite{Ali,Dai}.

The purpose of this paper is to first
provide in section 2,
a systematic discussion of the
general case where the latticized array is not a polygon and
moreover the vev's depend on the site location (the case of varying gauge
coupling in treated in the appendix). This will give rise,
in certain energy regimes, to a five-dimensional description in a
curved warped geometry instead of just the usual Kaluza--Klein case,
where the fifth dimensional is flat.
In section 3 we work out some examples completely in the
lattice formulation, whereas for some others we resort to the
continuum approximation.
We believe that these results can
be further used towards constructions that are more
phenomenologically oriented. Finally, in section 4, we review and
refine certain aspects
of the model introduced in \cite{sfe1} in connection with the more recent
developments. We also suggest a natural mechanism for the resolution
of singularities in solutions that appear in studies of the Coulomb
branch of the ${\cal N}=4$ SYM theory.
We have also written an appendix where we treat the case of varying gauge
coupling according to the lattice position (but constant vev). We show that
this is related to the case of varying vev's and constant gauge coupling
of section 2, by a generalization of the formulation of the usual
supersymmetric quantum mechanics, but now on the discrete lattice.

\section{The general formalism}

As in \cite{ACG1,HPW1} we start with an
effective four-dimensional low energy action
which is the sum of YM actions for $SU(k)$ as well as for a
Higgs field that provides the necessary linking
\be
S=-{1\ov 4 (N\!+\!1)g_4^2} \int d^4x  \sum_{j=0}^{N} \tr\bl( F_{\m\n}^j\br)^2
+ \sum_{j=0}^{N\!-\!1 }
\tr \bl((D_\m \Phi_{j})^\dagger D^\m\Phi_{j} \br)\ , \qq \m=0,1,2,3\ .
\label{acct}
\ee
Each gauge term is labeled by an integer $j$, whereas the Higgs field
$\Phi_{j}$ links the different gauge terms. In transforms in the
$({\bf N}_{j},{\bf \bar N}_{j+1})$ representation of
the gauge group, where we note
that group generators corresponding to different sites commute.
Accordingly the covariant derivative is defined as
$D_\m \Phi_{j}=\del_\m \Phi_{j}- i
A^j_{\m} \Phi_{j} + i \Phi_{j} A^{j+1}_{\m}$.
The gauge coupling $(N\!+\!1)g_4^2$ is
associated with each factor $SU(k)$ separately.
On the other hand the coupling $g_4^2$ is associated with the diagonal
$SU(k)$ that is left unbroken and dictates the physics at very low energies.
We assume that there is a potential for the Higgs field $\Phi_j$ that allows
this field to develop a vev $v_j= v f_j $,
where we find it convenient to separate an energy scale $v$ from a
set of numbers $f_j$ that characterize the strength of the vev at each site
$j$. This is our main difference from
\cite{ACG1,HPW1}, which however will have important consequences.
A similar action to \eqn{acct} has
appeared before in section 2 of \cite{FuNi2}, where in a five-dimensional
$U(1)$ gauge theory
on an anisotropic lattice, the continuum limit in the
four-dimensional layer was taken, but the fifth dimension was left latticized.

We digress, to note that,
the case with varying gauge coupling, and constant vev's are
completely related to the case of constant gauge coupling and varying
vev's that we examine here. This will be shown in the appendix.

Turning back to \eqn{acct} we see that the relevant part of the action
corresponding to the gauge field fluctuations is
\be
S=-{1\ov 4 (N\!+\!1) g_4^2  }\int d^4x  \sum_{j=0}^{N} \tr\bl( F^j_{\m\n}\br)^2
+ v^2 \sum_{j=0}^{N\!-\!1} f_j^2 \tr \bl(A^{j+1}_{\m}-A^j_{\m}\br)^2 \ .
\label{actt}
\ee
The mass term above can be written in terms of an $(N\!+\!1)\times(N\!+\!1)$
matrix $T$ defined by writing it in the form
$A_\m^t T A^\m$. Finding the mass spectrum amounts to simply diagonalizing
the matrix $T$. However, this can be a cumbersome procedure except
for a small number of sites.
In order to develop a systematic approach we prefer to write this mass
term as
\be
\sum_{j=0}^{N} A_{\m}^j T_j A^j_{\m}\ ,
\label{maza}
\ee
where the operator $T_j$ is defined as
\be
T_j = v^2 (e^{-d_j}-1) f_j^2
(e^{d_j}-1) = -4 v^2 \sinh(d_j/2) f_{j-\ha}^2 \sinh(d_j/2) \ ,
\label{tjt}
\ee
with $d_j$ being the usual differential operator satisfying $[d_j,j]=1$.
The rewriting \eqn{maza} of the mass term is correct up
to a term which is taken care of by the boundary conditions as discussed below.

In principle, in order to find the spectrum of the gauge field fluctuations we
need to diagonalize the operator $T_j$. This can be done
in some cases explicitly, but of course not in general.
Before we come to specific examples it is important to mention some generic
properties and features.
Consider the eigenvalue equation
\be
T_j \phi_{j}= M^2 \phi_{j}\ .
\label{ewj}
\ee
By construction the operator $T_j$ is Hermitian and therefore
it has real eigenvalues. In addition, since it can be written in the form
$T_j=q_j^\dagger q_j$, where $q_j=v f_j(e^{d_j}-1)$, its spectrum is
positive semi-definite , i.e. $M^2\geq 0$.
The eigenvalue equation \eqn{ewj} should be supplemented with
some boundary condition.
In order to find what consistent boundary condition can be imposed we
first realize that the equation
of motion for $A_\m^N$ as it follows with the mass term \eqn{maza} is
different than that following with a mass term as in \eqn{actt}, by the
extra term $v^2_N(A^{N\!+\!1}_\m-A^N_\m)$.
Translating this into a condition for
the eigenvectors $\phi_{j}$ we have
\be
{\rm boundary\ condition\ I}:\phantom{xxxx}
\phi_{N+1} = \phi_N\ .
\label{bb1}
\ee
This boundary condition preserves the zero mode.
An alternative boundary condition which does not preserve the zero mode is
\be
{\rm boundary\ condition\ II}:\phantom{xxxx}
\phi_0=0 \quad {\rm and\ (or)}\quad \phi_{N} =0\ .
\label{bb2}
\ee
The physical motivation and consistency for such a boundary condition
needs some further explanation. Let us relax the assumption that the
gauge coupling is the same at all sites and introduce a ``defect'' by
assuming that one of gauge couplings
at the site $j=N$ is much smaller than the couplings at the other
sites which are taken to be equal to $g_4^2$, i.e. $g_{4,N}^2\ll g_4^2$.
Hence, in this limit, after we scale $A^{N}_\m\to A^{N}_\m g_{4,N}$,
we see that the gauge field $A_\m^{N}$ drops out of the mass term in
\eqn{actt}. In terms of the corresponding eigenfunction $\phi_N$ this
implies the condition $\phi_N=0$ in \eqn{bb2}. A similar comment applies
if we take the gauge coupling at the site $j=0$ very small, which implies
then the boundary condition $\phi_0=0$.

Imposing the boundary condition \eqn{bb1} we find that there is one massless
eigenvalue corresponding to the zero model, as well as $N$ massive
ones. We emphasize that this procedure of finding the eigenvalues is
completely equivalent to simply diagonalizing the $(N\!+\!1)\times(N\!+\!1)$
matrix $T$ defined before by writing the mass term in \eqn{actt} in the form
$A_\m^t T A^\m$.
We can always construct a set of $N\!+\!1$
eigenvectors $\{\phi_{j,n},\ n=0,1,\dots, N\}$
of the operator $T_j$ that obey the eigenvalue equation \eqn{ewj}
as well as the orthogonality and completeness relations
\be
\sum_{j=0}^{N} \phi_{j,n} \phi_{j,m}^* = \d_{n,m}\ ,\qq
\sum_{n=0}^{N} \phi_{j,n} \phi_{k,n}^* = \d_{j,k}\ .
\label{grot}
\ee
The normalized wavefunction corresponding to the
zero mode is $\phi_{j,0} =(N+1)^{-1/2}$ with $M_0=0$.

On the other hand, imposing one of the boundary condition in
\eqn{bb2} we find that there are $N$ massive eigenvalues,
but the zero mode is projected out of the spectrum.
Then, in the finite
sums in \eqn{grot} we exclude the corresponding terms.

Lets us introduce a source term corresponding to a unit charge at the origin
of the coordinate system and at the $k$-th site.
Then, the differential equation that determines the potential energy at
site $j$
due to a charge located at site $k$ is given by
\be
\nabla_3^2 V_{j,k} - T_j V_{j,k} =
4\pi (N\!+\!1)g_4^2  \d_{j,k} {\d^{(3)}}({\bf x})\ .
\label{flteq}
\ee
We may expand $V_j$ is terms of the basis eigenvectors of the differential
operator $T_j$ as $V_{j,k}(r)=\sum_n \tilde V_{n,k}(r) \phi_{j,n}$.
Using \eqn{grot} we eventually arrive at
\be
V_{j,k}(r) =- {(N\!+\!1) g_4^2\ov r} \sum_{n=0}^{N} \phi_{j,n} \phi_{k,n}^*
e^{-M_n r}\ .
\label{jsf}
\ee

We may discuss quite generally,\footnote{For completeness, we note that,
when the eigenvectors
$\phi_{j,n}$ form a complete but not orthonormal set,
we may follow the procedure of Gram--Schmidt in order to
construct such a set. Otherwise, \eqn{grot} is replaced by
\be
\sum_{j=0}^{N} \phi_{j,n} \phi_{j,m}^* = A_{mn}\ ,\qq
\sum_{n,m=0}^{N} \phi_{j,n} A\inv_{nm} \phi_{k,m}^* = \d_{j,k}\ .
\label{grot2}
\ee
where $A_{mn}$ is some $(N\!+\!1)\times (N\!+\!1)$ Hermitian matrix.
Consequently, \eqn{jsf} is replaced by
\be
V_{j,k}(r) =- {(N\!+\!1)g_4^2 \ov r} \sum_{n,m=0}^{N} \phi_{j,n} A\inv_{nm}
\phi_{k,m}^* e^{-M_n r}\ .
\label{jsf2}
\ee
}
several features for
the behaviour of the potential in \eqn{jsf}. For instance,
at distances much larger that the typical size of the array, i.e. $r\gg N/v$,
obviously only the zero mode, if it exists, contributes
to the sum in \eqn{jsf}. Hence we have that
\be
{\rm Zero\ mode\ exists}: \phantom{xxx}
V_{j,k}(r)\simeq -{g_4^2\ov r}\ , \qq {\rm for} \quad r\gg N/v\ .
\label{cc1}
\ee
Therefore, at low energies we obtain the usual
Coulombic behaviour with coupling
constant $g_4^2$ (we ignore possible slow variations due to the running
of couplings).
This is the expected behaviour for a four-dimensional gauge theory for $SU(k)$
which is the unbroken
diagonal subgroup of the original $N+1$ copies of $SU(k)$.
This behaviour is universal for all sites due to the fact that the wavefunction
for the zero mode has the same value at all sites.
For the cases that the zero mode does not exist the typical behaviour is
\be
{\rm Zero\ mode\ does\ not\ exist}: \phantom{xxx}
V_{j,k}(r)\sim -{g^2_4 \phi_{j,1}\phi_{k,1}^* \ov r} e^{- M_1 r} \ ,
\qq {\rm for} \quad r\gg N/v\ ,
\label{cc1g}
\ee
where the inverse of the lowest eigenvalue, which is expected to be
comparable to the typical size of the array,
i.e. $M_1\inv \sim N/v$, sets the range of the Yukawa interaction.
Its effective overall
strength depends on the sites.\footnote{In \eqn{cc1g} the
strength
of the Yukawa interaction is constant. This is due to the discreteness of
the spectrum. In cases where the continuum limit is considered this strength
is $r$-dependent, giving rise to an overall factor for $e^{-M_1 r}$ which is
proportional to $1/r^{3/2}$ instead of $1/r$.}
In contrast, for distances $r\ll 1/v$ all modes in the sum in \eqn{jsf}
contribute equally. After using the completeness relation
in \eqn{grot} we obtain
\be
V_{j,k}(r)\simeq -{(N\!+\!1) g_4^2 \ov r} \d_{j,k}\ ,
\qq {\rm as} \quad r\ll 1/v\ .
\label{cc2}
\ee
This is the expected behaviour for a four-dimensional theory at high
energies where we have $N+1$ independent copies of gauge theories for
$SU(k)$, each with gauge coupling $(N\!+\!1)g_4^2$.
Since, the various copies are independent the potential is
non-vanishing only for charges charged with the same $SU(k)$
factor, i.e. $j=k$.
For charges charged with respect to different $SU(k)$
factors where $j\neq k$, the potential has an expansion in powers of
$r$ with leading term of order $g_4^2 v N$ for site-separation of order 1,
and $g_4^2 v/N$ for site-separation of order $N$. Hence, the magnitude
of the force between charges located at the same site is much larger
than that between charges located at different sites.
In the latter cases, higher loop diagrams that contribute may be considered
as, depending on the value of the coupling constant, they may give comparable
contributions (a similar comment has been made for a specific example
in \cite{ACG1}).

We conclude our general discussion
with further remarks on the general behaviour
of the potential as a function of $r$.
First, notice that, as it follows from \eqn{cc1}, we have implicitly chosen
opposite charges for the two charged particles with respect to the diagonal
$SU(k)$ subgroup that prevails at low energies.
The question is whether it is possible in some
models to have a repulsive instead of attractive force for some $vr\sim 1$,
although of course eventually the force has to turn attractive according to
\eqn{cc1}.
Certainly, the possibility of an repulsive force for some range
of $r$ is excluded for charges
at the same site since all terms in
\eqn{jsf} come with the same sign when $j=k$.
However, for $j\neq k$ not all terms have the same sign and
the possibility of a repulsive force is not excluded for some range of the
values for $r$. Nevertheless, in all of our examples in this paper the
force remains attractive for all $r$, in agreement with physical expectations.

\subsection{Continuum limit and a fifth dimension}

A quite interesting behaviour occurs for intermediate distances, which
are much larger that the lattice spacing, but also much smaller than its
typical size $\sim N/v$, namely for $1/v \ll r \ll N/v$. Here we assume that
$N\gg 1$.
Then we may approximate the lattice by a
continuum with fifth space-variable $x_5=j/v$.
Then the differential
operator $T_j$ is well approximated by
$T(x_5)= -\del_5 f(x_5)^2 \del_5$,
where $f(x_5)$ is the continuum limit of the
sequence of numbers $\{f_j\}$ which is assumed to exist.
Then, the continuous limit of the action \eqn{actt} is written as
\be
S= -{1\ov 4 g_5^2} \int d^4x dx_5   \tr\left( F_{\m\n}^2 +
f(x_5)^2 \bl(\del_5 A_{\m}\br)^2 \right) \ ,\qq g_5^2=N g_4^2/v\ .
\label{actt3}
\ee
This can be reinterpreted as the five-dimensional action
$-{1\ov 4 g_5^2} \int d^5x \sqrt{-G} \tr\bl( F_{MN}^2\br)$,
where $M=(\m,5)$ and in the gauge $A_5=0$, defined
in the background with a metric given by the line element
\be
ds^2 = f(x_5)^2 \eta_{\m\n} dx^\m dx^\n + dx_5^2 \ .
\ee

Hence, what we have derived is that there is an energy regime
where we may safely use a five-dimensional gauge theory defined in a
curved, in general, warped spacetime,
in order to compute physical processes in a
gauge theory in the usual four-dimensional Minkowski
spacetime. The four-dimensional gauge theory is renormalizable whereas
the five-dimensional one is not and only emerges as an
effective description in the appropriate energy regime.

We make the important note that the regime where $1\ll vr\ll N$ is a
strict continuum limit where the operator $T_j$ is well approximated by
the second order differential operator $T(x_5)=-\del_5 f(x_5)^2 \del_5$.
Higher terms that appear in the derivative expansion of
$T_j$ are neglected.
However, there is a less restrictive continuum limit that retains all such
higher derivative terms.
As it will also be confirmed in the examples below, this limit is attained
for $v r\ll N$ and therefore is valid for distances $vr\simeq 1$, where
the approximation that leads to \eqn{actt3} fails.

For future convenience it is better to use the variable $z$ instead of $x_5$,
with change of variables as $dx_5=e^A dz$. Also we will
rename $f^2$ by $e^{2 A}$.
The equation determining the potential between two charges is still given by
\eqn{flteq} with $T_j$ replaced by $T(x_5)$.
The corresponding eigenvalue equation for this operator becomes
\be
T \phi_n = -e^{-A} \del_z e^A \del_z \phi_n(z)= M^2_n \phi_n(z)\ ,
\label{uu2}
\ee
where we have used the variable $z$ instead of $x_5$. The orthogonality
and completeness relations read
\be
\int dz e^A  \phi_{n}(z) \phi^*_{m}(z) = \d_{n,m}\ ,\qq
\sum_{n=0} \phi_{n}(z) \phi^*_{n}(z') =  e^{-A} \d (z\!-\! z')\ .
\label{grrot}
\ee
As in the discrete case, we may distinguish the following two cases:
If the spectrum of the operator $T$ has a mass gap
then the behaviour of the potential between charges is, for large
distances, of the Yukawa-type, whereas if there is
no such mass-gap we have a power-law type of behaviour.

We note that, even though for notational purposes we have used the discrete
set of number $\{n\in Z\}$ in order to label our states, the spectrum can be
discrete of continuous depending on the function $A(z)$. In order to check
that it is convenient to cast the differential equation in \eqn{uu2} as
a Schr\"odinger differential equation. Indeed, after defining $\phi_n =
e^{-A/2} \Psi_n$ we obtain
\be
-\del_z^2 \Psi_n + V_{\rm Sch.} \Psi_n = M^2_n \Psi_n\ ,\qq
V_{\rm Sch.}={1\ov 4} (\del_z A)^2 +\ha \del^2_z A\ .
\label{ppoo}
\ee
This potential has the form of the potentials appearing in supersymmetric
quantum mechanics \cite{SQMwit}
with corresponding superpotential $W=-\ha \del_zA$.
Hence, the spectrum is positive semi-definite as already
expected from the fact that
it originates from the spectrum of the operator $T_j$ which we have already
argued that it is positive semi-definite.

If the Schr\"ondinger potential in \eqn{ppoo} can be neglected for
some wide range of values
of the variable $z$, then in this range the wavefunctions are approximated
by plane-waves as $\Psi_{k}\simeq 1/\sqrt{2 \pi} e^{i k z}$. Then, using
that $\phi_k=e^{-A/2} \Psi_k =f(x_5)^{-1/2} \Psi_k$ we obtain that
the potential between charges at positions $z$ and $z'$ in the fifth
dimension, is given by
\be
V_{z,z'}(r) \simeq -{g_5^2 \ov \pi}\
{e^{-\ha(A(z)+A(z')}\ov r^2 +(z\!-\! z')^2}\ .
\label{ldb}
\ee
This behaviour resembles the usual Kaluza--Klein case, but with an effective
coupling that depends on the fifth dimension. This is due to the fact that the
effective fifth dimension that emerges is not flat.
Such behaviour will be encountered in some of the explicit examples below.

\section{Solvable examples}

The examples that we specifically work out in this section belong to
two categories: Those where we work directly in the lattice formulation
and those where the continuous approximation based on \eqn{actt3} is employed
from the very beginning. We first consider the case
of a periodic array on a polygon that has been considered before in this
context in \cite{ACG1,HPW1} and its supersymmetric
variant already within the context of
the AdS/CFT correspondence in \cite{sfe1}.
In our second example the periodicity condition is relaxed, resulting into
a model that has a discrete mass spectrum, but no zero mode.
Both cases share the feature that the vev's are the same at all sites.
Our third example has vev's that vary with the site position, but even so
it turns out that is is completely solvable with the help of
Laguerre polynomials.
Finally, within the continuous approximation we consider two classes of
examples with varying vev's. Both have a continuous spectrum but one of them
has a mass gap.

\subsection{Periodic array on a polygon}

In this example we give the same vev at all sites, i.e. $v_j=v$.
Moreover, because of the $Z_{N+1}$ symmetry of the array,
we require periodicity of the wave function as $\phi_{j+N+1}=
\phi_{j}$.
Since we have that $f_j=1$ the operator in \eqn{tjt} is simply
$T_j=-4 v^2 \sinh^2(d_j/2)$. This commutes with the differential operator
$d_j$ and therefore they have the same set of
eigenfunctions which when properly
normalized read
\be
\phi_{j,n}={1\ov \sqrt{N\!+\!1} }
e^{2\pi i {jn\ov N+1}}\ , \qq j=0,1,\dots ,N
\ , \quad n=0,1,\dots , N\ .
\label{eei1}
\ee
The $N+1$ mass eigenvalues are given by
\be
M_n = 2 v \sin\left(\pi n\ov N\!+\!1\right)\ , \qq n=0,1,\dots , N\ ,
\label{hwgd}
\ee
where we also observe that a zero mode exists.
Since $|\phi_{j,n}|=1/\sqrt{N\!+\!1}$
the potential \eqn{jsf} is the same for all
sites, as expected due to the $Z_{N\!+\!1}$ symmetry of the array.
Even in this simple case, it does not seem possible to
perform the sum in \eqn{jsf} explicitly.\footnote{However,
we may easily prove that for $r\ll 1/v$, the behaviour of the effective
potential is
$V_{j,k}\simeq -\d_{j,k} g_4^2(N\!+\!1)/r -{\pi v g^2_4/(N\!+\!1)}
\left(\sin^2\left(\pi(j-k)/(N\!+\!1)\right)
-\left(\ha\pi/(N\!+\!1)\right)^2\right)^{-1}+ \cdots$. This is in
accordance with the general behaviour that
we have stated after eq. \eqn{cc2}.}
Nevertheless, the
behaviour for $r\gg N/v$ and $r\ll 1/v$ is that predicted on general
grounds by \eqn{cc1} and \eqn{cc2}, respectively.

The continuum limit holds for $ v r \ll N$ and $j\!-\! k\ll N$,
Then the original discrete periodic array is replaced by a continuum
periodic array, but we retain all terms in the derivative
expansion of the operator $T_j$. The expression
for the potential turns out to be given by the following integral
\be
V_{j,k}(r)=-2  {g_4^2 N\ov \pi r} \int_{0}^{\pi/2} d\varphi
\cos \left(2(j\!-\! k)\varphi\right) e^{-2 v r \sin\varphi}\ .
\label{jklp}
\ee
We may evaluate this
in terms of generalized hypergeometric functions, but
we will refrain from presenting the result since
the resulting expressions are quite complicated.

If we further restrict to distances $1\ll vr\ll N$ we find the behaviour
\be
V_{j,k}(r)
\simeq -{g_5^2\ov \pi}
{1\ov r^2 + (x_5\!-\! x'_5)^2}\ ,\qq {\rm for } \quad 1/v \ll r \ll N/v\ ,
\label{jh1}
\ee
where $x_5=j/v$ and $x'_5=k/v$.
This is the regime where the description in terms of a five-dimensional
theory in flat space (since $f(x_5)=1$) becomes appropriate.
Then the continuum periodic array is replaced by a an infinite continuum array.
In that case the
properly normalized
wavefunctions are $\phi_k(x_5)=e^{i k x_5}/\sqrt{2\pi }$, with
$z$ taking values in the whole real line, whereas the mass spectrum $M^2=k^2$
is continuous. Then using that $V=-g_5^2/r  \int_{-\infty}^\infty
dk e^{-|k| r}/(2\pi )$ we exactly reproduce \eqn{jh1}. This is the typical
Kaluza--Klein behaviour corresponding to the five-dimensional effective
theory \eqn{actt3} and was already found in \cite{ACG1,HPW1}.

There is in addition the scale regime $vr\gg 1$ where the original discrete
periodic array is replaced by a infinite, but nevertheless discrete array.
Then the potential is given by an infinite sum which can be
computed exactly
\ba
V_{j,k}(r) &=& -{g_4^2\ov r} \sum_{n=-\infty}^\infty e^{2\pi i (j-k)n/N}
e^{-2 \pi v r |n|/N}   =  - g_5^2 \sum_{n=-\infty}^\infty {1\ov
r^2 + (j\!-\! k - N n )^2/v^2}
\nonumber\\
& = & -{g_4^2\ov r}\ {\sinh (2\pi r v/N)\ov
\cosh(2\pi r v/N)-\cos\left(2\pi (j\!-\! k)/N\right)}\ .
\label{jf9}
\ea
This expression interpolates between those in \eqn{jh1} for
$1\ll vr\ll N$ and \eqn{cc1} for $vr \gg N$.
In the former case the corrections to \eqn{jh1} are in powers of $vr/N$,
whereas in the latter case the corrections to \eqn{cc1} are Yukawa-like,
corresponding to the Kaluza--Klein modes of a reduction to a circle of
radius $N/v$.

\subsection{Relaxing the periodicity condition}

As in the previous example we give the same vev to all sites,
but we no longer impose the periodicity condition. Instead, we impose
the boundary condition \eqn{bb2}
that the wave function $\phi_{j}$
vanish at the endpoints, namely that $\phi_{0}=\phi_{N}=0$. The properly
normalized wavefunctions that form a complete set are
\be
\phi_{j,n}= \sqrt{2\ov N} \sin(\pi j n/N)\ ,\qq
j=0,1,\dots , N\ ,\quad n=1,2,\dots , N\!-\!1\ .
\ee
The mass spectrum is given by
\be
M_n=2 v \sin\left(\pi n\ov 2N\right)\ , \qq n=1,2,\dots , N\!-\!1\ ,
\ee
where the zero mode is projected out.
With these eigenfunctions and eigenvalues the sum
in \eqn{jsf} cannot be computed exactly.
Nevertheless, we may show that the
behaviour for $r\gg N/v$ and $r\ll 1/v$ is that predicted on general
grounds by \eqn{cc1} and \eqn{cc2}, respectively.

In the limit $vr\gg 1$ and for $N\gg 1$ we have that (we set $L=N/v$)
\ba
V_{j,k}(r) & = & -{2 g_4^2\ov  r } \sum_{n=1}^{\infty}
\sin (\pi j n/ N) \sin(\pi k n/N) e^{-\pi v n r/N}
\nonumber\\
& = & -{g_4^2\ov r}
{\sin \left(\pi j\ov N\right) \sin \left(\pi k\ov N\right)
\sinh\left(\pi r\ov L\right) \ov
\left[\cosh\left(\pi vr\ov N \right) - \cos\left(\pi (j+k)\ov N\right)\right]
\left[\cosh\left(\pi vr\ov N \right) - \cos\left(\pi (j-k)\ov N\right)\right]
}\ .
\ea
For $r\ll x_5\sim x'_5 \ll L$, where $x_5=j/v$ and $x'_5=k/v$, we obtain
the same behaviour as in \eqn{jh1}.
Also
\be
V_{x_5,x'_5}
\simeq -{4 g^2_4 L x_5 x'_5 \ov \pi r^4}\ , \qq x_5,\ x'_5 \ll r \ll L\ ,
\ee
which is a behaviour similar to that of a seven-dimensional theory.
Finally, for distances much larger than the length $L$ we have
\be
V_{j,k}\simeq
- {2 g_4^2\ov r} \sin(\pi j/N)  \sin(\pi k/N)e^{-\pi r/L} \ ,
\qq {\rm for } \quad r\gg L\ ,
\ee
in accordance with the general formulae \eqn{cc1}.

There is also a continuum limit for distances $rv\ll N$,
that goes beyond the two derivative approximation \eqn{cc1g}
\be
V_{j,k}(r)= -2 {g_4^2 N\ov \pi r} \int_0^{\pi/2} d\varphi
\sin(2 j\varphi)\sin(2 k\varphi)\ e^{-2 v r \sin\varphi}\ .
\ee
Again this is expressible in terms of generalized hypergeometric functions.

\subsection{Array with varying vev's}

Cases with vev's varying according to the site position are in general
more difficult to handle for increasing site number $N$.
However, let us consider the case of
$v_j=v \sqrt{j}$, $j\geq 1$.\footnote{For convenience, we have shifted the
range of $j$ by one. Hence, $j$ starts from $j=1$ and not from $j=0$ as in the
general formulation before. We take this into account in the rest of this
subsection.}
Though not obvious,
it is possible to
completely determine in general
the eigenvalues and eigenfunctions of the operator
$T_j$ which now equals
\be
T_j = v^2 (e^{-d_j}-1) j ( e^{d_j}-1)\ .
\label{opqpe}
\ee
As in some quantum mechanical problems it is convenient to pass
to the momentum representation where the operator $d_j$ is just a
number denoted by $i p$ and the site position $j$ becomes the differential
operator $i \del_p$.
Then the eigenfunctions in the momentum
representation satisfy the differential equation
\be
i v^2 (e^{i p}-1) \del_p\left( (e^{i p}-1) \phi_p\right) = M^2 \phi_p\ ,
\ee
with properly normalized solution\footnote{We note that the case of an array
with $v_j=v j$, $j\geq 1$ is also solvable by applying the same idea.
Indeed, in this case after we substitute $\phi_p=(1-q)^\m F(q)$, where
$q=\cos^2(p/2)$ and $\m={1/4} +i/2 \sqrt{m^2-1/4}$,
we obtain that $F(q)$ satisfies a hypergeometric equation
with $a=b=\m$ and $c=1/2$ in the standard notation.
We also note that, normalizability of $\phi_p$ requires that the mass
spectrum has a gap, $M_{\rm gap}=v/2$.}
\ba
&& \phi_p = {i\ov \sqrt{\pi}} { e^{i m^2/2 \cot(p/2)}\ov e^{i p}-1}=
{1\ov 2\sqrt{\pi} i } e^{im^2/2 \cot(p/2)} (1+ i \cot(p/2))\  ,
\nonumber\\
&& \phantom{xxxxxxxxxxxxxxxxxxxxxxxx}
0\leq  p \leq 2 \pi\ ,\quad m = {M/v}\ .
\ea
In order to compute the potential between charges we need the
wavefunction in the position representation $\phi_j$. This is given by
\be
\phi_j = \sqrt{1\ov \pi} \int_0^{2 \pi} dp \sin(j p) \phi_p\ .
\label{l2p2}
\ee
After changing variables as $x=\cot(p/2)$ we have to compute the integral
in \eqn{l2p2} which takes the form
\be
\phi_j  =  -{1\ov 2\pi } \int_{-\infty}^\infty dx
\left( {(x+i)^{j-1}\ov (x-i)^{j+1}} - {(x-i)^{j-1}\ov (x+i)^{j+1}}\right)
e^{i m^2 x/2 } (1+ i x)\ .
\label{hd}
\ee
The integrand has poles at $x=\pm i$ in the complex plane and we
may compute it using standard methods in complex analysis. Since $m^2>0$
we may choose the contour of integration to close in the upper half plane
and to include $x=i$.
Then, only the first term in the integrand in \eqn{hd} contributes.
The result is
\ba
\phi_j & = & {1\ov (j-1)!} {\del^{j-1}\ov \del x^{j-1}}
\left[ e^{-m^2 x/2} (x+1)^{j-1}\right]_{x=1}
\nonumber\\
& = &  e^{-m^2/2} \sum_{n=0}^{j-1} {(j-1)! (-m^2)^n\ov (j-1-n)! (n!)^2} \ .
\label{kjla}
\ea
One recognizes the last sum as an expression for the
Laguerre polynomials $L_{j-1}(m^2)$. Therefore
\be
\phi_j=  e^{-m^2/2} L_{j-1}(m^2)\ ,\qq j=1,2,\dots \ .
\label{laagg}
\ee
Having obtained this solution in a systematic way, it is important to
verify directly that \eqn{laagg} indeed satisfies the eigenvalue equation
$T_j\phi_j=M^2 \phi_j$, where $T_j$ is given by \eqn{opqpe}.
This is quite straightforward, using the recursion
relations for the Laguerre polynomials, $L_{j-1}(m^2)$.

\subsubsection{The limit of infinite array}

In the case that the array is large, so that $N\gg 1$,
boundary effects can be neglected for $vr \ll N$.
Then the
mass spectrum is continuous with no mass gap ($M^2\geq 0$).
Using the orthogonality
and completeness properties of the Laguerre polynomials for $m^2\in (0,\infty)$
we see that \eqn{grot} are satisfied.
The potential between charges at sites $j$ and $k$ is then
\be
V_{j,k}(r)= -2{g_4^2 N\ov r}\int_0^\infty dm m e^{-m^2 - m vr} L_{j-1}(m^2)
L_{k-1}(m^2)\ ,\qq {\rm for} \quad vr\ll N\ .
\label{pooer}
\ee
For specific values for $j$ and $k$ we may give the result in terms of the
error function ${\rm erf}(vr/2)$ and powers of $vr$,
but we will not present any results explicitly.

It is easy to see that for $vr\ll 1$ we obtain \eqn{cc2} whereas for $vr\gg 1$
and small values for $j$ and $k$ we obtain an $1/r^3$ fall-off for the
potential as
\be
V_{j,k}(r)\simeq -2 {g_4^2 N\ov v^2 r^3}\ ,\qq {\rm for} \quad
j\!-\! k  \ll vr\ll N\ .
\ee
This is characteristic of the behaviour of a six-dimensional theory.

The limit of site position $j\gg 1$ and $m^2\ll 1$ is also
of particular interest.
Then using properties of the Laguerre polynomials we find that
\be
\phi_j \simeq J_0(2 m \sqrt{j})\ , \qq {\rm as} \quad j\gg 1\ , \phantom{x}
m^2\ll 1 \phantom{x} {\rm and} \phantom{x} j m^2={\rm finite}\ ,
\label{hhew}
\ee
where $J_0$ is the zeroth order Bessel function. This is the result that one
would have obtained from the five-dimensional two-derivative
effective theory based on \eqn{actt3}, as we will shortly verify
(see footnote 6 below).

\subsubsection{Finite arrays}

In the case that the array is finite with $j=1,2,\dots , N+1$
we have to consider boundary effects.
There are two choices, one projecting out the zero mode and one preserving it.

\no
\underline{Projecting out the zero mode}: We impose the boundary condition
\eqn{bb2} that the wavefunction vanishes at the
end, namely that $\phi_{N+1} = 0$.
This condition determines the allowed values for $m^2_n$ as the $N$
zeros of the
corresponding Laguerre polynomial $L_{N}$
\be
L_N(m_n^2)= 0 \ , \qq n=1,2,\dots , N\ .
\ee
The resulting $N$ eigenfunctions
$\phi_{j,n}=e^{-m^2_n/2} L_{j-1}(m^2_n) ,\ j=1,2,\dots , N$ with $n=
1,2,\dots , N$ form a complete orthogonal set, but there are no longer
properly normalized to 1.
Using the Darboux--Christoffel formula for the Laguerre polynomials
we find that the correctly normalized orthogonal complete set of eigenvectors
is given by
\be
\phi_{j,n}={m_n\ov N |L_{N\!-\!1}(m_n^2)|} L_{j-1}(m_n^2)\ ,
\qq j=1,2,\dots , N, \quad n=1,2,\dots , N\ .
\label{hgfj1}
\ee
Then the potential between charges located at the
lattice-sites $j$ and $k$, and separated by a distance $r$ is
\be
V_{j,k}(r)= -{g_4^2 (N+1)\ov r}\sum_{n=1}^{N} {m_n^2\ov N^2
L_{N\!-\!1}^2(m_n^2)}\ L_{j-1}(m_n^2) L_{k-1}(m_n^2) e^{-m_n v r}\ .
\label{pie2}
\ee

As an example consider the simplest non-trivial case with $N=2$.
The two roots of
$L_2(m^2)=0$ are given by $m_\mp^2=2\mp\sqrt{2}$.
Using the fact that $L_0(x)=1$ and $L_1(x)=1-x$ we find that the
two eigenfunctions forming a complete orthonormal set
according to \eqn{grot}, are
\be
\phi_{1,\mp}  = {m_\pm\ov 2}\ ,\qq \phi_{2,\mp}  = \pm {m_{\mp}\ov 2} \ .
\ee
The potential between charges
contains of course only Yukawa-type terms. We have
\be
V_{ij}(r) = -{g_4^2 \ov r} (a_{ij}^- e^{-m_-v r} + a_{ij}^+ e^{-m_+ v r})\ ,
\qq i,j=1,2\ ,
\label{gh7}
\ee
where the $2\times 2$ matrices ${\bf a^\mp}$ are given by
\be
{\bf a^\mp} = {3\ov 4} \pmatrix{2\pm\sqrt{2} & \pm \sqrt{2}\cr
\pm \sqrt{2} & 2\mp \sqrt{2}}\ .
\ee

\no
\underline{Keeping the zero mode}: Let us now
impose the boundary condition \eqn{bb1},
that is $\phi_{N+2}=\phi_{N+1}$,
or using the properties of the Laguerre polynomials, $\phi'_{N+2}=0$.
That determines $N$ nonzero mass eigenvalues $m_n^2$, $n=1,2,\dots ,N$
from the condition
\be
L'_{N+1}(m_n^2)=0\quad  \Rightarrow \quad L_{N+1}(m_n^2)=L_N(m_n^2)=
L'_N(m_n^2) \ ,
\qq n=1,2,\dots , N\ .
\ee
However, there is also the zero mode which
is preserved by the boundary condition.
As before, using the
Darboux--Christoffel formula for the Laguerre polynomials
we find that the correctly normalized orthogonal complete set of eigenvectors
is given by
\ba
\phi_{j,n}& =& {1\ov \sqrt{N\!+\!1} |L_{N}(m_n^2)|}\ L_{j-1}(m_n^2)\ , \qq
j=1,2,\dots , N\!+\!1\ ,\quad n=1,2,\dots , N\ ,
\nonumber\\
\phi_{j,0} & =&  {1\ov \sqrt{N\!+\!1}}\ ,\qq j=1,2,\dots , N+1\ .
\ea
Then the potential between charges at sites $j$ and $k$ has a Coulombic
as well as Yukawa terms
\be
V_{j,k}= -{g_4^2\ov r} -{g_4^2\ov r}
 \sum_{n=1}^N {1\ov L^2_N(m_n^2)}\ L_{j-1}(m_n^2)
L_{k-1}(m_n^2) e^{-m_n vr}\ .
\ee

As an example consider again the case of $N=2$. Now there are three
mass eigenvalues given by $m_0^2=0$ and the two zeroes of the algebraic
equation $L'_3(m^2)=0$ which are $m_\mp^2=3\mp\sqrt{3}$.
The corresponding eigenfunctions are
\be
\phi_{1,\mp}={\sqrt{3}\pm 1\ov 2\sqrt{3}}\ ,\quad
\phi_{2,\mp}={\pm 1 -\sqrt{3}\ov 2\sqrt{3}}\ ,\quad
\phi_{3,\mp}={\mp} {1\ov \sqrt{3}}\ ,\quad
\phi_{j,0}={1\ov \sqrt{3}}\ .
\ee
The potential between charges is
\be
V_{j,k}(r)= -{g_4^2\ov r} - {g_4^2\ov r}
(b_{jk}^- e^{-m_- vr} + b_{jk}^+ e^{-m_+ vr} )\ ,\qq j,k=1,2,3\ ,
\label{gh8}
\ee
where the $3\times 3$ matrices ${\bf b}^\mp$ are given by
\be
{\bf b^\mp} = {\ha} \pmatrix{2\pm\sqrt{3} & -1 & -1\mp \sqrt{3}\cr
 -1& 2\mp \sqrt{3} & \pm \sqrt{3}-1 \cr
-1\mp\sqrt{3}& \pm \sqrt{3} -1& 2}\ .
\label{bpm}
\ee
%

Finally, we note that when $N$ becomes very large,
the spectrum in both cases that we have considered becomes continuous
and the above expressions approach
the corresponding ones in \eqn{laagg} and \eqn{pooer}.
This can be seen using the property of the Laguerre polynomials
\be
L_n(x)= {e^{x/2}\ov \sqrt{\pi} (nx)^{1/4}} \cos(2 \sqrt{nx}-\pi/4)
+{\cal O}(n^{-3/4})\ ,\qq {\rm for}\quad n\gg 1\ .
\ee

\subsection{Example within the continuous approximation}

We consider now cases where the lattice is already treated in the
continuum approximation where a description in terms of a gauge theory
in a five-dimensional warped spacetime \eqn{actt3}
is valid, namely when $1\ll vr\ll N$.
The distribution of vev's might be such that the mass parameter
that sets the scale at these distances is not $v$ itself,
but some other mass parameter $M_0$.
This is assumed to be such that $v/N\ll M_0\ll v$ which implies that the
corresponding length scale $1/M_0$ is within the allowed distances for the
continuum approximation to be valid.

First, let us consider the case with
$f(x_5)=2 M_0 x_5$, $x_5\geq 0$. Applying the general formalism we easily find
that the variable $x_5$ and the variable $z$ that enters in \eqn{uu2}
are related by
$2 M_0 x_5=e^{2 M_0 z}$, with $-\infty < z < \infty$ and that
the function $A(z) = 2 M_0 z $. Therefore the Schr\"odinger
potential in \eqn{ppoo} is just a constant, i.e. $V_{\rm Sch.}=M_0^2$.
That defines a mass gap since in order to have normalizable wavefunctions
we must require that $M^2\ge M_0^2$. We obtain
\be
\phi_M(z) = {e^{-M_0 z}\ov 2 \sqrt{\pi}}
(M^2-M_0^2)^{-1/4} e^{ i (M^2\!-\! M_0^2)^{1/2} z}\ .
\ee
Hence, the potential between charges at positions $z$ and
$z'$ in the fifth dimension is
\be
V_{z,z'}=  -{g_{5}^2 M_0 e^{-M_0(z+z')}\ov \pi R}
K_1(M_0 R)\ ,\qq R=\sqrt{r^2+(z\!-\! z')^2}\ ,
\ee
where $K_1$ is the modified Bessel function of order one.
If we consider distances
much smaller than the characteristic scale of the mass gap $1/M_0$,
then we obtain indeed the behaviour \eqn{ldb}.
Instead, in the opposite limit of distances much larger than $1/M_0$
we find that
\be
V_{z,z'}(r) \simeq -{g_5^2\ov \sqrt{2 \pi}}
e^{-M_0(z+z')} {M_0^2\ov (M_0 R)^{3/2}}
e^{-M_0 R}\ ,
\qq {\rm for } \quad 1/M_0 \ll R \ll N/v \ .
\ee
Naturally, the range of Yukawa interaction is set by the mass gap, whereas
the strength has the $1/r^{3/2}$ behaviour as noted in footnote 2.

Also,
the value of the mass gap $M_0$ is in agreement with what was noted
in footnote 5 for the exact lattice treatment of the vev distribution
$v_j=v j$, after we set $M_0=v/2$.

\subsection{Another example within the continuous approximation}

A generalization of the previous case which however does not lead to a mass
gap for the spectrum, is to take
$f(x_5)=\left( 2(\n+1) M_0 x_5\right)^{2 \n+1\ov 2(\n+1)}$  with $x_5\geq 0$
and where the exponent of $x_5$ is introduced as such for later notational
convenience. We find that the variables $x_5$ and $z$ are related as
$2(\n+1) M_0 x_5= (M_0 z)^{2(\n+1)}$ and also that $e^A =(M_0 z)^{2\n+1}$.
Hence, the Scrh\"odinger potential takes the form
\be
V_{\rm Sch.} = {\n^2 -1/4\ov z^2} \ ,
\ee
from which we conclude that the mass spectrum is continuous without a mass
gap.
The eigenfunctions are given in terms of Bessel functions as
\be
\phi_M(z) = {1\ov \sqrt{2 M_0}} (M_0 z)^{-\n} J_\n(M z)\ .
\label{jd}
\ee
It turns out that\footnote{Note that, for $M_0= v/2$ and $\n=0$ we have that
$v_j=v \sqrt{j}$. Then indeed the wavefunction \eqn{jd}
agrees with \eqn{hhew}.}
the potential between charges at the same position $z$
in the fifth dimension is computed in terms of a hypergeometric function
as (for simplicity we do not deal with the case of charges at different
positions in the fifth dimension)
\ba
&& V_{z,z}(r)=-{2 \G(\n+3/2)\ov \sqrt{\pi} \G(\n+1)}  {g_5^2/M_0^{2\n+1}
\ov  r^2 (r^2+4 z^2)^{\n+1/2}} F\left(\n+1/2,\n-1/2,2\n+1,{4 z^2\ov
r^2+ 4 z^2}\right)\ ,
\nonumber\\
&& \phantom{xxxxxxxxxxxxxxxxxxxxxxxxxxxxxxxxxxxx}
{\rm for} \quad 1/v\ll r\ll N/v\ .
\ea
We also note the limiting case
\be
V_{z,z}(r)= -{2 \G(\n+3/2)\ov \sqrt{\pi} \G(\n+1) } \
{g_{5}^2/M_0^{2\n+1} \ov r^{2\n+3}}\ ,\qq {\rm for} \quad 1/v \ll z\ll r\
\ee
and that for $r\ll z$ we confirm the behaviour \eqn{ldb}.


\section{Strong coupling considerations}

We would like to study similar effects at strong coupling using the AdS/CFT
correspondence \cite{malda}. We will see that our findings and the emergence
of an effective fifth dimension are in precise agreement with the
gauge theory considerations so far.

The clearest discussion for our purposes can be done
for the
Coulomb branch of the ${\cal N}=4$ SYM theory with gauge group $SU(Nk)$.
This theory has six scalars $X^I$, $I=1,2,\dots ,6$
in the adjoint of the gauge group.
We will reanalyze and refine the model introduced in \cite{sfe1}
and we will study the potential between a quark and an antiquark.
We give vev's to two of the six scalars in the theory in a way that the
original $SO(6)$ global invariance of the theory that rotates the six
scalars, breaks to $SO(4)\times Z_N$.
We may represent these scalars as $Nk\times Nk $ traceless matrices.
In general, since the scalar potential in the SYM action is proportional to
$\sum_{I<J}\tr[X^I,X^J]^2$ we can choose the vev's along the diagonal
so that the potential is zero. That preserves sixteen supercharges and
breaks the conformal invariance.
We arrange the $Nk\times Nk$ matrices $X^I$'s in the form of
block-diagonal matrices with $N$ diagonal
$k\times k$ blocks, where in each block
the vev's are the same.
The ${\cal N}=4$ SYM theory action contains the term
$\sum_{I=1}^6 {\rm tr}(D_\m\Phi^I)^2$ which couples
the gauge fields and the six scalars. It is convenient to choose the standard
real basis for the $SU(Nk)$ generators
$J_{ij}= e_{ij} -{1\ov Nk} \d_{ij} I_{Nk\times Nk}$,
where the matrix elements of the matrices
$e_{ij}$ are $(e_{ij})_{mn}=\d_{im}\d_{jn}$.
Hence, we give vev's to the scalars represented by the six-dimensional
vector $\vec \Phi$, as $\vec \Phi = h_{i} \vec X^i_{\rm vev}$,
where $h_{i}=J_{ii}$ are the diagonal Cartan generators.
The masses of the gauge fields arise from the term
$\cL_{\rm mass}\sim
\sum_{I=1}^6 {\rm tr} [\Phi^I,A_\m]^2$. After we define $A_{\m}=A_\m^{ij}
J_{ij}$ we have
\be
\cL_{\rm mass} \sim k \sum_{i,j=0}^{N-1} (\vec X^i_{\rm vev}\! \cdot
\vec X^j_{\rm vev} - \vec X^i_{\rm vev}\! \cdot \vec X^i_{\rm vev})
A_{\m}^{ij} A^{\m,ji}\ .
\label{jk2}
\ee

This basic formula can now be applied to specific cases.
As in \cite{sfe1} we choose for the vev's in the $j$th block the values
\be
\vec X^j_{\rm vev}
=(0,0,0,0,r_0 \cos(2\pi j/N),r_0 \sin(2\pi j/N))\ ,\qq j=0,1,\dots ,
N\!-\! 1\ .
\label{doois}
\ee
These form an $N$-sided polygon enclosed by a circle in the
$x_5$-$x_6$ plane of radius $r_0$.
Then we find that \eqn{jk2} becomes
\be
\cL_{\rm mass}\sim r_0^2 k \sum_{i,j=0}^{N\!-\!1}
\sin^2(\pi (j-i)/N) A_{\m}^{ij} A^{\m,ji}\ .
\label{jk24}
\ee
From this we read off the masses as
\be
M_n \sim r_0 \sqrt{k} \sin(\pi n/N)\ , \qq n=0,1,\dots , N\!-\!1\ ,
\label{kj2}
\ee
which are the same as the ones we found in \eqn{hwgd}. They have a
degeneracy $d_n$ which is: $d_0=N\!-\!1$ and $d_n=2(N\!-\!n)$, for $n\geq 1$
(Note that $\sum_{n=0}^{N\!-\!1}d_n=N^2\! -\! 1$). Using the fact that
$M_n=M_{N-n}$ it is easily seen that the expression for the potential is
given by $V=-{g_4^2/(N r)} \sum_{n=0}^{N\!-\!1} d_n  e^{-M_n r}=
-{g^2_4/r}\left(\sum_{n=0}^{N\!-\!1} e^{-M_n r} -1/N\right)$,
which, for large enough $N$, is the same as the potential \eqn{jsf}
corresponding to the mass formulae in \eqn{hwgd}
(for test particles at the same lattice cite).
That makes apparent that the model in \cite{sfe1} and the construction
of \cite{ACG1,HPW1} (reviewed in subsection 3.1)
describe essentially the same Physics.

The advantage of having a D3-brane realization in a model with an emerging
latticed fifth dimension, is that we may consider
the theory at strong coupling using the AdS/CFT correspondence.
In the type-IIB theory
we consider a large number $Nk$ of D3-branes that are distributed in the
transverse to the branes six-dimensional space according to \eqn{doois}.
Namely, the gravitational metric that we will consider is
\be
ds^2 = H^{-1/2} \eta_{\m\n} dy^\m dy^\n + H^{1/2} (dx_1^2+\cdots + dx_6^2)\ ,
\label{metrhg}
\ee
where $y^\m$, $\m=0,1,2,3$ are the parallel to the brane directions.
The harmonic function $H$ in the six-dimensional space spanned
by $x_i$, $i=1,2,\dots , 6$ is given by
\be
H= \sum_{j=0}^{N\!-\!1} {R_4^4\ov |\vec x- \vec X^j_{\rm vev} |^4} \ ,
\ee
where $R_4^4=4\pi g_4^2 k$ is the 't Hooft coupling associated with the
$SU(k)$ gauge theory. We see that
the vev's in the field theory \eqn{doois} become the centers in the
harmonic function above.
We emphasize that the supergravity approximation is valid even close to each
one of the centers if we choose that $R_4^4\gg 1$ and also small values
for $g_4^2$ which implies that there are many branes located at each
center, i.e. $k\gg 1$. The number of such centers $N$ need not be large.

The harmonic function $H$ has been computed in \cite{sfe1}. It takes the
form
\be
H = R^4 {U^2+r_0^2\ov \left( (U^2+r_0^2)^2-4 r_0^2 u^2\right )^{3/2}} \S_N\ ,
\label{hf29}
\ee
where $R^4=4 \pi g_4^2 k N$. Also
$U^2 = x_1^2+ \dots + x_6^2$, $u^2= x_5^2 + x_6^2$ and
\be
\S_N = {\sinh N x\ov \cosh N x - \cos N \psi} + N {
\left( (U^2+r_0^2)^2-4 r_0^2 u^2\right )^{1/2}\ov U^2 +r_0^2}
{\cosh N x \cos N \psi - 1\ov (\cosh N x - \cos N \psi)^2} \ ,
\ee
where $\psi$ is the polar angle in the $x_5$-$x_6$ plane and
\be
e^x = {U^2 +r_0^2\ov 2 r_0 u} + \sqrt{ \left(U^2+r_0^2\ov 2 r_0 u\right)^2 -1}
\ .
\ee
Notice that $H$ explicitly processes the symmetry $SO(4)$ in the
four-dimensional subspace spanned by $x_1,x_2,x_3$ and $x_4$ as well as
the $Z_N$ symmetry under the shift $\psi\to \psi +2\pi/N$.

The potential between a pair of heavy quark and antiquark
separated by a distance $L$ is found\footnote{We denote the
distance between the
charges by $L$ in order to confirm with standard notation in the AdS/CFT
literature. This is nothing but the distance denoted by $r$ in the previous
part of the paper.}
by minimizing the Nambu--Goto action of a fundamental string in the
supergravity background \eqn{metrhg}. We will not give details as the
necessary techniques we first developed for the conformal case in
\cite{marey}.
For backgrounds of the type that we have here we will use eqs. (5)-(7)
in \cite{brand1}.
The trajectory that minimizes the action lies on the $x_5$-$x_6$ plane where
$u=U$. However, not every angle $\psi$ is allowed, but only those with
$\sin (N\psi)=0$. All angles satisfying this condition are equivalent because
of the $Z_N$ symmetry. For concreteness we may choose the
trajectory along the $x_5$-axis (for $\psi =0$ and $u=U$).
Let's define the function $f=R^4/H$. Then for the trajectory that we are
interested in we have
\ba
f(U)& =& {(U^2-r_0^2)^3\ov U^2+r_0^2} {1\over \S}\ ,
\nonumber\\
\S & =&  {(U/r_0)^N+1\ov (U/r_0)^N-1} +
{2 N \ov (U/r_0)^N+(r_0/U)^N-2}{(U/r_0)^2-1\ov (U/r_0)^2+1}  \ ,
\ea
where we have used the fact that $e^x=U/r_0$, when $u=U$ and $\psi=0$.
Then the integrals for the length $L$ and the energy $V$ are given
by
\be
L \ =\ 2 R^2 \int_{U_0}^\infty
dU \sqrt{f(U_0)\ov f(U)(f(U)-f(U_0))} \ ,
\label{le1}
\ee
and
\be
V\ = \
\frac{1}{\pi} \int_{U_0}^\infty dU \left[ {\sqrt{ f(U)\ov
f(U) - f(U_0)}} - 1 \right] - {1\ov \pi} (U-r_0) \ ,
\label{en1}
\ee
where $U_0$ denotes the lowest value of $U$ that can be reached by the
string geodesic.
In principle, we have to evaluate the integrals and try to express
$V$ as a function of $L$. This is impossible in general though it is
easy to deduce that $V$ is a monotonously increasing functions of
$L$ from $-\infty$ to $0$ in the interval $L\in (0,\infty)$.

\subsection{Examining various energy regimes}

As usual, in the AdS/CFT correspondence large (small) distances in the
five-dimensional geometry, corresponds
to small (large) distances in the gauge theory side \cite{wisu}.
The variable $U$ represents a distance in the AdS side and it
is interpreted as an energy variable in the gauge theory side.
It is instructive to find the various limits of the metric \eqn{metrhg}
and of the quark-antiquark potential,
by examining the behaviour of the harmonic function $H$
as we go from the UV to the IR.
The UV corresponds to energies large enough so that the vev's can be neglected.
The deep IR corresponds to energies very close to the vev's.

\subsubsection{Behaviour for $U\gg r_0$}

In the UV for $U\gg r_0$ since the harmonic takes the form
\be
H\simeq {R^4 \ov U^4}\ , \qq {\rm for}\quad U\gg r_0 \ ,
\ee
the metric becomes that for $AdS_5\times S^5$
with each factor having radius $R$ in string units,
Then the original $SO(6)$ global symmetry of the metric is restored.
That corresponds to the fact that in the deep UV the values of the vev's can
be neglected and we should be describing an $SU(Nk)$ SYM theory.
In this regime the integrals can be evaluated and give
for the potential energy between a quark and antiquark the
results of \cite{marey}. Hence we have
\be
V\simeq
-  2 \left({\G(3/4)\ov \G(1/4)}\right)^2 {R_4^2 \sqrt{N} \ov L}\ ,
\qq L\ll {R^2_4 \sqrt{N}\ov r_0}\ .
\label{lsq6}
\ee
This behaviour is similar to the one found from
pure gauge theory considerations in \eqn{cc2} for the interaction
of charges at the same site $j=k$.

\subsubsection{Behaviour for $U-r_0\gg r_0/N$}

For distances from the location of the centers
much larger than the distance between the different centers
we may approximate the harmonic function by \cite{sfe1}
\be
H = R^4 {U^2+r_0^2\ov \left( (U^2+r_0^2)^2-4 r_0^2 u^2\right )^{3/2}}
\ , \qq {\rm for}\quad U-r_0\gg r_0/N\ ,
\label{halm}
\ee
since then $\S_N\simeq 1$.
Then we see that the discrete part $Z_N$ of
the symmetry group $SO(4)\times Z_N$ becomes a continuous $U(1)$.
Then the resulting integrals for the length and the energy
are quite complicated (see eqs. (A.6) and (A.7) of \cite{brand1})
and it does not seem possible to compute them
explicitly, let alone
to explicitly obtain the energy as a function of the separation length.
This approximation is valid for $L\ll R_4^2 N/r_0$.
The situation here is similar to the one described from a gauge theory
point of view by eq. \eqn{jklp} (for $j=k$).

As a separate remark we note that the metric \eqn{metrhg} with the above
harmonic \eqn{halm} has a ring singularity defined by the equations
$U^2=u^2=x_5^2+x_6^2=r_0^2$. It is obvious that this is
an artifact of the continuous approximation which is not valid near the
singularity, where the harmonic which should be used is given by \eqn{halm1}
below and the metric then is completely non-singular.
The same resolution of singularities applies to solutions that have been
used for studies of the Coulomb branch of the ${\cal N}=4$ SYM gauge theory
in, for instance, \cite{brand1,bakas1,fgpw2}.
These solutions, take the form \eqn{metrhg}, but unlike \eqn{hf29},
the distribution of D3-branes is continuous instead of discrete.
It is obvious that the solutions will become non-singular if we replace
the continuous by a discrete distribution of D3-branes with the given
continuous limit. An example of that was already given in the appendix
of \cite{sfe1} representing the discrete version of a continuous distribution
of D3-branes on a disc.

As another side remark we note that the solution with harmonic \eqn{hf29}
as well as its continuous approximation \eqn{halm}, do not have a
five-dimensional gauged supergravity origin.

\subsubsection{Behaviour for $r_0/N \ll U-r_0\ll r_0$}

For distances from the centers
much larger than the distance between the different centers,
but also much smaller than the radius of the circle,
we may approximated the harmonic function by
\be
H\simeq {R^4\ov 4r_0 U^3}\ , \qq {\rm for}\quad  r_0/N \ll U-r_0\ll r_0\ .
\ee
In this case we have a smeared D3-brane solution along one transverse
direction which is T-dual to a D4-brane solution. Also the global
symmetry has been enhanced from an $SO(4)\times U(1)$ into an $SO(5)$.
Since a D4-brane solution corresponds to a five-dimensional theory
we expect that there will be a $1/L^2$ behaviour for the quark-antiquark
potential.
Indeed we obtain the result \cite{brand1}
\be
V\simeq
-4\sqrt{\pi} \left(\G(2/3)\ov \G(1/6)\right)^3 {R_4^4 N \ov r_0 L^2}\ ,
\qq {\rm for}\quad {R^2_4 \sqrt{N}\ov r_0} \ll  L\ll {R^2_4 N\ov r_0}\ .
\label{1l2}
\ee
The behaviour is similar to \eqn{jh1} from the usual
gauge theory point of view.

\subsubsection{Behaviour for $U-r_0\ll r_0/N$}

In the deep IR for energy scales very close to the vev's
\be
H\simeq {R_4^2\ov U^4} \ , \qq {\rm for}\quad  U-r_0\ll r_0/N\ ,
\label{halm1}
\ee
where $R^4_4= 4\pi g_4 k$ is the 't Hooft coupling for an $SU(k)$ gauge theory.
Then we have
\be
V\simeq
-  2 \left({\G(3/4)\ov \G(1/4)}\right)^2 {R_4^2 \ov L}\ ,
\qq {\rm for}\quad L\gg {R^2_4 N\ov r_0}\ .
\label{lsq}
\ee
This is a Coulombic behaviour for a gauge group $SU(k)$
and is similar to that in \eqn{cc1} from the usual gauge theory point of view.

\subsubsection{Discrete infinite array}

Finally, we
may also find a regime where the original circular discrete array is
approximated by a discrete infinite array on a straight line. This is
possible for distances from the centers much smaller that the radius of
the circle. From the gauge theory point of view at weak coupling the
analogous potential
is described by \eqn{jf9}.
The solution has a metric given by \eqn{metrhg} with harmonic
\ba
H & = &
\sum_{j=-\infty}^\infty {R_4^4\ov \left(U^2+ (x_6 -2\pi v j)^2\right)^2}
\nonumber\\
& = & {R_4^4\ov 4 U^3 v^2 }\left({v \sinh(U/v)\ov \cosh(U/v)-\cos(x_6/v)} + U
{\cosh(U/v) \cos(x_6/v)-1\ov (\cosh(U/v)-\cos(x_6/v))^2}\right) \ ,
\ea
where $U^2=x_1^2+x_2^2+\dots + x_5^2$ and $v=r_0/N$.
Translated into separation distances for the quark-antiquark pair, this
approximation is valid when $L\gg R_4^2 \sqrt{N}/r_0$.
The trajectory that we will use to compute the potential amounts to setting
$x_6=0$. Then we utilize the general formulae \eqn{le1} and \eqn{en1}
with $R$ replaced by $R_4$ and the function $f$ being given by
\be
f(U)= 4 v U^3\left(\coth\left(U/(2 v)\right) +{{U/(2 v)}
\ov \sinh^2 \left(U/(2 v)\right)}\right)^{-1}
\label{gfg}
\ .
\ee
As we move from $U\gg v$ to $U\ll v$ this case interpolates between those
described before by \eqn{1l2} and \eqn{lsq}.
It is not possible to compute the energy as a function of the separation
explicitly, but we may use the above expression in order to find the
first correction to the $1/L^2$ behaviour in \eqn{1l2}. Expanding
\eqn{gfg} for $U\gg v$ and keeping the two leading terms we find that
\be
V= - 4\sqrt{\pi} \left(\G(2/3)\ov \G(1/6)\right)^3
{R_4^4\ov v L^2} \left(1+ c_1
{R_4^6\ov v^3 L^3} e^{-c_2 {R_4^4\ov v^2 L^2}} + \cdots
\right)\ ,
\label{coorr}
\ee
where the $c_1$ and $c_2$ are two constants of oder 1.\footnote{Precisely,
$c_1 = \sqrt{8\pi^3}\left({8\ov 3 \sqrt{6}}-1\right)
\left(\G(2/3)\ov \G(1/6)\right)^2\simeq 0.083$ and
$c_2= 4\pi \left(\G(2/3)\ov \G(1/6)\right)^2\simeq 0.744 $.}
Since $1\ll {R^4_4\ov v^2 L^2}\ll N$ the corrections to the leading
$1/L^2$ behaviour are small. However, they do not obey a power-like law as the
corrections to \eqn{jh1} (noted after \eqn{jsf}).
Similarly, the corrections to \eqn{lsq} are found by expanding \eqn{gfg} for
$U\ll v$.  It is easy to see that these corrections
are organized in powers of ${R_4^2 N\ov r_0 L}\ll 1$,
in contrast to the corrections to \eqn{cc1}
which are of the Yukawa type (noted again below \eqn{jsf}).
Hence, it seems that
the mechanisms for developing an effective fifth dimension are
different at weak and strong coupling.

In will be interesting to further extend the results of this paper in order to
incorporate fermions and to investigate the renormalization of
the gauge coupling constant. Also an extension to non-commutative gauge
theories could be of some interest.

\bs\bs
\centerline{\bf Acknowledgements}

I would like to thank A. Kehagias for a discussion.
This research was supported
by the European Union under contracts
TMR-ERBFMRX-CT96-0045 and -0090, by the Swiss Office for Education and
Science, by the Swiss National Foundation and by the contract
HPRN-CT-2000-00122.

\appendix
\section{Varying gauge coupling}
\setcounter{equation}{0}
\renewcommand{\theequation}{\thesection.\arabic{equation}}

In this appendix we first study differential operators defined on the
lattice of the type
\be
T_j = q^\dagger_j q_j \ , \qq j=0,1,\dots ,N\ ,
\label{lpq2}
\ee
for some operator $q_j$ and
which have energy eigenvalues that are
positive semi-definite, i.e. $M^2\geq 0$.
In the case of the operator \eqn{tjt} discussed in the main text we have
that $q_j=v f_j (e^{d_j}-1)$, as already noted.
If a state $\phi_{j,0}$ satisfies the equation $q_j \phi_{j,0}=0$, then
automatically we have a solution of $T_j\phi_{j,0}=0$ corresponding to a zero
mass eigenvalue. Define now the lattice differential operator $\tilde T_j$
obtained
by reversing the order of $q_j$ and $q_j^\dagger$ in \eqn{lpq2}
\be
\tilde T_j = q_j q^\dagger_j \ , \qq j=0,1,\dots ,N\ .
\label{lp45}
\ee
The discussion next is completely parallel to that in
supersymmetric quantum mechanics (SQM) \cite{SQMwit}.
The spectra and eigenfunctions of $M_n^2$ and $\tilde M_n^2$ of
$T_j$ and $\tilde T_j$
\ba
&&\tilde M^2_n = M^2_{n+1}\ ,\quad M^2_0=0\ ,
\nonumber\\
&& \tilde \phi_{j,n}= {1\ov M_{n+1}} q_j \phi_{j,n+1}\ ,
\label{hdlp}\\
&& \phi_{j,n+1}= {1\ov \tilde M_{n}} q^\dagger_j \tilde \phi_{j,n}\ .
\nonumber
\ea
Hence, the spectra of the operators $T_j$ and $\tilde T_j$
are identical except for the zero mode. As we have seen, boundary
conditions may
project it out of the spectrum in which case the spectra are identical.
Moreover, acting with $q_j$
($q^\dagger_j$) we may convert an eigenvector of $T_j$ ($\tilde T_j$) into
an eigenvector of $\tilde T_j$ ($T_j$).

Let us define the matrices
\be
H_j=\pmatrix{T_j & 0 \cr 0 & \tilde T_j} \ ,\quad
Q_j=\pmatrix{0 & 0 \cr q_j & 0}\ , \quad
Q^\dagger_j=\pmatrix{0 & q^\dagger_j \cr 0 &  0}\ .
\ee
It is trivial that these obey the closed superalgbebra
\be
[H_j,Q_j]=[H_j,Q_j^\dagger]=\{Q_j,Q_j\}=\{Q_j^\dagger,Q_j^\dagger\}=0 \ ,\quad
\{Q_j,Q_j^\dagger\}=H_j\ .
\ee
In terms of this superalgebra, the fact that
$Q_j$ and $Q^\dagger_j$ commute with $H_j$ is responsible for
the degeneracy of the spectra.

As an application of the above general results, let us consider again the
low energy effective action in the lattice \eqn{acct}, but now with varying
gauge coupling $g_{4,j}^2$ and Higgs values $v_j$.
The relevant term derived from \eqn{acct} corresponding to
the gauge field fluctuations is
\be
S=-{1\ov 4 (N\!+\!1) }\int d^4x  \sum_{j=0}^{N} \tr\bl( F^j_{\m\n}\br)^2
+ \sum_{j=0}^{N\!-\!1}
v_j^2 \tr \bl(g_{4,j+1} A^{j+1}_{\m}-g_{4,j} A^j_{\m}\br)^2 \ .
\label{actt9}
\ee
Consider first the case of constant vev's $v_j=v$.
The mass term above can be written as
\be
\sum_{j=0}^{N} A_{\m}^j \tilde T_j A^j_{\m}\ ,
\label{maza9}
\ee
where the operator $\tilde T_j$ is defined as
\be
\tilde T_j = v^2 g_{4,j} (e^{-d_j}-1)
(e^{d_j}-1) g_{4,j} = -4 v^2 g_{4,j} \sinh^2(d_j/2) g_{4,j}\ .
\label{tjt9}
\ee
Hence, the operator $\tilde T_j$ has precisely the form \eqn{lp45}
with $q_j=g_{4,j}(e^{-d_j}-1)$. Using the formalism that we developed
we may map this problem into a problem for the superpartner
operator $T_j$ in \eqn{lpq2}. Comparing with the case of constant gauge
coupling and varying vev's considered in section 2, we find that they
are identical after we identify $v^2 g^2_{4,j+\ha}$ by $v^2 f^2_{j-\ha}$.

As a specific application consider the case of couplings
$g_{4,j}=g_4 \sqrt{j}$, $j\geq 1$. After using the results of subsection 3.3
we find that the set of eigenfunctions of the operator $T_j$ are
\be
\phi_{j} \sim  L_{j}(m^2)\ ,\qq m={M\ov g_4 v}\ .
\ee
Using \eqn{hdlp} we find that
the eigenfunctions of the operator \eqn{tjt9} are
\be
\tilde \phi_{j} \sim  \sqrt{j}
\left(L_{j-1}(m^2) - L_j(m^2)\right)\ .
\ee
We may easily verify, using recursion relations for the Laguerre polynomials,
that this solution satisfies the eigenvalue
equation for the operator $\tilde T_j$ in \eqn{tjt9}.
We have left undetermined the overall normalization constant, since this
would depend on the boundary conditions that we will impose.
Note that a choice of a boundary condition for either
$\phi_j$ or $\tilde \phi_j$ implies the boundary condition for the other,
if we want that the mass spectra are also related according to \eqn{hdlp}.
Of course we may still impose independent boundary conditions in which case
the two spectra are not related.

Finally, we note that in the most general
case of varying vev's and gauge couplings,
the operator $T_j$ still takes the form \eqn{lpq2} with
$q_j=v_j (e^{d_j}-1)g_{4,j}$.



\begin{thebibliography}{99}
\bibitem{Kim}
H.~D.~Kim,
{\it To be (finite) or not to be,
that is the question "Kaluza-Klein contribution to the Higgs mass},
{\tt hep-ph/0106072}.
\end{thebibliography}
\end{document}